\documentclass[12pt]{article}
\usepackage{qPartitionFuns}
\usepackage{cancel}
\input{MyQcircuit.sty}
\begin{document}

\title{Use of Quantum Sampling to Calculate\\
 Mean Values of Observables and Partition Function\\
 of a Quantum System}

\author{Robert R. Tucci\\
        P.O. Box 226\\
        Bedford,  MA   01730\\
        tucci@ar-tiste.com}

\date{ \today}

\maketitle

\vskip2cm
\section*{Abstract}
We describe an algorithm for
using a quantum
computer to calculate
mean values of observables
and the partition
function of a quantum system.
Our algorithm
includes two sub-algorithms.
The first sub-algorithm
is for calculating,
with polynomial efficiency,
certain diagonal matrix
elements of an observable.
This sub-algorithm is performed
on a quantum computer,
using quantum phase estimation
and tomography.
The second sub-algorithm
is for sampling a probability
distribution. This
sub-algorithm is not
polynomially efficient.
It can be performed either
on a classical or a quantum
computer, but
a quantum computer
can perform it quadratically
faster.

\newpage

\section{Introduction}
In this paper,
we describe an algorithm for
using a quantum
computer to calculate
mean values of observables
and the partition
function of a quantum system.
Our algorithm includes
two sub-algorithms.

One sub-algorithm
is for calculating,
with polynomial efficiency,
certain diagonal matrix
elements of an observable.
This sub-algorithm is performed
on a quantum computer,
using quantum phase estimation
and tomography.
This sub-algorithm is very similar to
the algorithm of Ref.\cite{Har}
by Harrow
et al.,
except that we modify
it to accomplish a
substantially different job
that has nothing to do whatsoever
with systems of linear equations.

A second sub-algorithm
is for sampling a probability
distribution. This
sub-algorithm is not
polynomially efficient.
It can be performed either
on a classical or a quantum
computer. However,
a quantum computer
can perform it quadratically
faster that a classical computer,
if one uses a
quantum sampling
technique based on
Szegedy operators, like, for instance,
the quantum Gibbs sampling
algorithm described in Ref.\cite{TucGibbsSam}.

We end the paper
with a brief section
comparing the algorithms
proposed in this
paper with
quantum algorithms proposed
in earlier papers for
calculating the same things.
In particular, we
compare our work to Ref.\cite{Woc1} by
Wocjan et al., Ref.\cite{Woc2}
by Poulin and Wocjan,
and Ref.\cite{Tem} by Temme et al..

\section{\cancel{Location},
\cancel{Location}, \cancel{Location}\\
Notation, Notation, Notation}

In this section, we will
define some notation that is
used throughout this paper.
For additional information about my
notational quirks, I recommend that
the reader
consult
the notation
section of some of
my previous papers; for example,
Ref.\cite{TucMetHas}.

We will often
use the symbol $\nb$ for the number ($\geq 1$) of qubits and
$\ns = 2^\nb$ for the number of states with $\nb$ qubits.
The quantum computing literature
often uses $n$ for $\nb$ and $N$
for $\ns$, but we will avoid this
notation. We prefer to use $n$
for the number operator $\ket{1}\bra{1}$.

Let $Bool =\{0, 1\}$. As usual,
let $\ZZ, \RR, \CC$ represent the set
of integers (negative  and non-negative),
real numbers, and
complex numbers, respectively.
We will also sometimes add a superscript
to the symbols
$\ZZ, \RR$ to indicate
a subset of these sets.
 For example, we will use $\RR^{\geq 0}$
to denote the non-negative reals.
For integers $a$, $b$
 such that $a\leq b$, let
$Z_{a,b}=\{a, a+1,
\ldots b-1, b\}$.

We will use $\Theta(S)$
to represent the ``truth function";
$\Theta(S)$ equals 1 if statement $S$ is true
and 0 if $S$ is false.
For example, the Kronecker delta
function is defined by
$\delta^y_x=\delta(x,y) = \Theta(x=y)$.

If $\vec{x} = x_{\nb-1} \ldots x_2 x_1 x_0$,
where $x_\mu\in Bool$, then
$dec(\vec{x}) = \sum^{\nb-1}_{\mu=0} 2^\mu x_\mu=x$.
Conversely, $\vec{x}=bin(x)$.
However, when
our meaning is clear from context,
we will omit the $bin()$ and $dec()$.
Hence, in some places $x$ might
stand for an element of $Z_{0,\ns-1}$,
and in other places
for the corresponding element of $Bool^\nb$.

We won't usually
put a caret over a symbol
to indicate that it is an
operator, but sometimes we will.
For example, we will
usually use $H$ for a Hamiltonian,
but sometimes, for clarity,
we will call it $\hat{H}$.

Note that in our quantum circuit diagrams,
time flows from the right to the left
of the diagram (this is the {\bf Dirac Convention}).
 Careful:
Many workers in Quantum
Computing draw their diagrams
so that time flows from
left to right (the {\bf Quayle Convention}).

We will say a problem
can be solved {\bf with
polynomial efficiency}, or {\bf
p-efficiently} for short, if
its solution can be
achieved in a
time polynomial in $\nb$. Here $\nb$ is
the number of bits
required to encode the
input for the algorithm
that solves the problem.

By {\bf compiling a unitary matrix},
we mean decomposing it into a {\bf SEO}
(Sequence of
Elementary Operators),
where by
elementary operators we mean
operators that act on only a few
qubits (usually 1, 2 or 3), such as
single-qubit rotations and CNOTs.
Compilations can be either
exact, or approximate (within
a certain precision).

We will say a unitary operator $U$
acting on $\CC^\ns$ can be
{\bf compiled with polynomial efficiently},
or {\bf p-compiled} for short,
if $U$
can be expressed,
either approximately
or exactly,
as a SEO
of length polynomial in $\nb$.
When necessary,
we specify whether
a p-compilation
is exact or approximate.

Next, we
explain
our notation related
to Discrete Fourier Transforms.

For any $x\in Z_{0,\ns-1}$,
define $k_x$ by

\beq
k_x = \frac{2\pi x}{N_S}
\;.
\eeq
Let $\hat{k}$
and $\hat{x}$
be operators acting on $\CC^\ns$
with eigenvectors and eigenvalues
given by
\beq
\hat{x}\ket{x} = x\ket{x}
\;
\eeq
and

\beq
\hat{k}\ket{\hat{k}=k_x} = k_x\ket{\hat{k}=k_x}
\;
\eeq
for any $x\in Z_{0,\ns-1}$.
Let the eigenvectors
of $\hat{x}$ and $\hat{k}$
be related by
a ``Discrete Fourier Transformation":

\beq
\ket{\hat{k}=k_x}=
\frac{1}{\sqrt{N_S}}\sum_{y=0}^{N_S-1}
e^{ik_x y}\ket{y}
\;.
\label{eq-dft}
\eeq
Eq.(\ref{eq-dft}) defines
a ``basis-changer"
unitary operator $U_{FT}$ with
matrix elements given by

\beq
\bra{y}U_{FT}\ket{x}=
\av{y|\hat{k}=k_x}
=\frac{1}{\sqrt{N_S}}e^{ik_x y}
=\frac{1}{\sqrt{N_S}}e^{ix k_y}
\;.
\eeq

Assume that the eigenstates
of $\hat{x}$ are orthonormal
and complete:

\beq
\av{y|x}= \delta(y,x)
\;
\eeq
for all $x,y\in Z_{0,\ns-1}$,
and

\beq
\sum_{x=0}^{\ns-1} \ket{x}\bra{x}=1
\;.
\eeq
Then it follows
that the eigenstates of $\hat{k}$
are orthonormal
and complete too, because

\beq
\av{\hat{k}=k_y|\hat{k}=k_x}
=\bra{y}U^\dagger_{FT} U_{FT}\ket{x}
=\delta(y,x)
\;,
\eeq
and

\beq
\sum_x \ket{\hat{k}=k_x}\bra{\hat{k}=k_x}
=\sum_x U_{FT}
\ket{x}\bra{x}
U_{FT}^\dagger
=1
\;.
\eeq

Often in quantum computing
we come across quantum states of the form

\beq
\ket{\hat{k}=k} =
\frac{1}{\sqrt{N_S}}\sum_{y=0}^{N_S-1}
e^{ik y}\ket{y}
\;,
\eeq
with
\beq
k = k_z + \Delta k
\;,
\label{eq-k-def}
\eeq
for some $z\in Z_{0,\ns-2}$
and $0\leq \Delta k \leq \frac{2\pi}{\ns}$.
If $k$ is neither
$k_z$ nor $k_{z+1}$
but lies somewhere in between, then
 $\ket{k}$ is close to but not
 exactly equal to an eigenstate of $\hat{k}$.
Note that when $\hat{k}=k_z$,

\beq
\bra{x}U_{FT}^\dagger\ket{\hat{k}=k_z}
=
\av{x|z}
=
\delta(x,z)
\;.
\eeq
For $\hat{k}=k$ where $k$ is given
by Eq.(\ref{eq-k-def}), this generalizes to

\beqa
\bra{x}U_{FT}^\dagger\ket{\hat{k}=k}
&=&
\frac{1}{N_S}\sum_{y=0}^{N_S-1}
e^{i(k-k_x) y}
\\&=&
\frac{1}{N_S}
\frac{1-e^{i(k-k_x)\ns}}
{1-e^{i(k-k_x)}}
\\&=&
\frac{1}{N_S}
\frac{e^{i(k-k_x)\frac{\ns}{2}}}
{e^{i(k-k_x)\frac{1}{2}}}
\frac{\sin\left((k-k_x)\frac{\ns}{2}\right)}
{\sin\left((k-k_x)\frac{1}{2}\right)}
\;.
\eeqa

\section{Diagonal Matrix Elements}
In this section,
we give an algorithm for calculating certain
diagonal matrix elements p-efficiently,
using a quantum computer. This algorithm
will be used as a subroutine in the algorithms
proposed in later
sections.

The algorithm described
in the proof of Claim \ref{cl-diag-mat-elem}
below is very similar to
the algorithm of Ref.\cite{Har},
except that we modify
it to accomplish a different job.
In a nutshell, Ref.\cite{Har}
combines two operations that were
separately familiar to
most quantum computerists
long before Ref.\cite{Har}: a phase estimation (PE)
operation,
followed by a quantum tomography operation.

 \begin{figure}[h]
    \begin{center}
    \epsfig{file=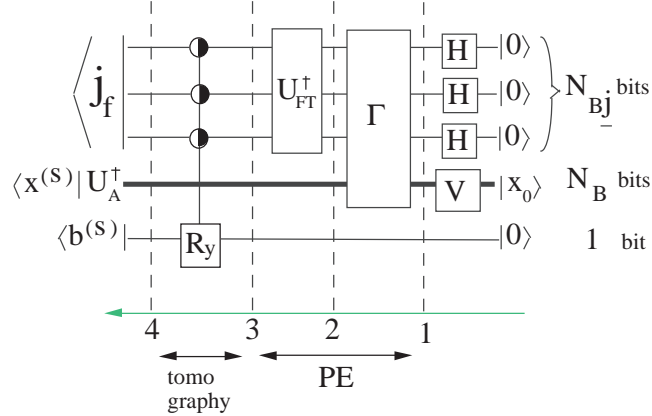, height=2.2in}
    \caption{Quantum circuit used in
    Claim \ref{cl-diag-mat-elem} to calculate
    the diagonal matrix element
    given by Eq.(\ref{eq-tar-diag-mat-elem}).}
    \label{fig-tom-pe}
    \end{center}
\end{figure}

\begin{claim}\label{cl-diag-mat-elem}
Let $A$ be an $\ns$ dimensional Hermitian matrix
with non-negative eigenvalues,
$V$ an $\ns$ dimensional unitary matrix,
$f:\RR^{\geq 0}\rarrow \RR^{\geq 0}$,
and $x_0\in Bool^\nb$, where $\ns=2^\nb$.
Assume that $f$ is simple (that is, that
it can be calculated p-efficiently).
Assume that we know how to p-compile $V$ and
$e^{iA\Delta t}$ for any $\Delta t \geq 0$.
Then we can calculate p-efficiently the
diagonal matrix element

\beq
\mu(x_0)=
\bra{x_0}V^\dagger f(A) V \ket{x_0}
\;.
\label{eq-tar-diag-mat-elem}
\eeq
\end{claim}
\proof

Let
$\nsj=2^\nbj$
be the number of states
of $\nbj$ bits, for a
set of bits labeled $\rvj$.

The diagonal matrix element
Eq.(\ref{eq-tar-diag-mat-elem})
can be calculated by
running on a quantum
computer the quantum circuit
shown in Fig.\ref{fig-tom-pe}.
In that figure, there
are three sets of qubits.
At the top are $\nbj$ ``probe" qubits.
Below the probe qubits are the
$\nb$ ``main" qubits
which are initially in state $\ket{x_0}$.
Finally, below the main
qubits is a single ``ancilla"
qubit that is used in the final tomography step.

In a moment,
we will describe
the evolution of
the state vector
as it courses down
this quantum circuit.
But before doing so,
we need to specify some
of the blocks in Fig.\ref{fig-tom-pe}
more precisely.

Note that
\beq
H^{\otimes \nbj}\ket{0}^{\otimes \nbj} =
\frac{1}{\sqrt{\nsj}}\sum_{j=0}^{\nsj-1}\ket{j}
\;.
\eeq
As for the $\Gamma$
box, it represents
\beqa
\begin{array}{c}
\Qcircuit @C=1em @R=2em @!R{
&\multigate{3}{\mbox{$\Gamma$}}
&\qw
\\
&\ghost{3}{}
&\qw
\\
&\ghost{3}{}
&\qw
\\
&\ghost{3}{}
&\qw
}
\end{array}
&=&
\begin{array}{c}
\Qcircuit @C=1em @R=.5em @!R{
&\dotgate
&\qw
&\qw
&\qw
\\
&\qw
&\dotgate
&\qw
&\qw
\\
&\qw
&\qw
&\dotgate
&\qw
\\
&\gate{U_{PE}^{2^{\nbj-1}}}\qwx[-3]
&\gate{U_{PE}^{2^1}}\qwx[-2]
&\gate{U_{PE}^{2^0}}\qwx[-1]
&\qw
}
\end{array}
\label{eq-pe-3bits}
\\
&=&
\sum_{j=0}^{\nsj-1}
\begin{array}{c}
\Qcircuit @C=1em @R=2em @!R{
&\gate{\ket{j}\bra{j}}
&\qw
\\
&\gate{U_{PE}^j}\qwx[-1]
&\qw
}
\end{array}
\;,
\eeqa
where

\beq
U_{PE}= e^{i A \Delta t}
\;.
\eeq
(For definiteness, some parts of
Eq.(\ref{eq-pe-3bits})
assume that
$\nbj=3$.)
Finally, the operation
with the ``half moon" nodes represents

\beqa
\begin{array}{c}
\Qcircuit @C=1em @R=.5em @!R{
&\muxorgate
&\qw
\\
&\muxorgate
&\qw
\\
&\muxorgate
&\qw
\\
&\qw
&\qw
\\
&\gate{R_y}\qwx[-4]
&\qw
}
\end{array}
&=&
\sum_{j=0}^{\nsj-1}
\begin{array}{c}
\Qcircuit @C=1em @R=1em @!R{
&\gate{\ket{j}\bra{j}}
&\qw
\\
&\qw
&\qw
\\
&\gate{R_j}
\qwx[-2]
&\qw
}
\end{array}
\;,
\label{eq-muxor}
\eeqa
(note that some parts of Eq.(\ref{eq-muxor})
again assume that $\nbj=3$.)
where $R_j$ is defined by

\beq
R_j = \left[
\begin{array}{cc}
c_j&-s_j\\
s_j& c_j
\end{array}
\right]
\;
\eeq
with

\beq
c_j = \sqrt{\gamma f(\frac{2\pi j}{\Delta t \nsj})}
\;,
\eeq

\beq
s_j = \sqrt{1-c_j^2}
\;.
\eeq
$\gamma$ is defined so that $0\leq c_j \leq 1$
for all $j\in Z_{0,\nsj-1}$.
Eq.(\ref{eq-muxor}) is an SU(2) multiplexor
with $\nbj$ controls. It can
be p-compiled approximately (the
software application
``Multiplexor Expander"\cite{qusann}
can do this).

Let

\beq
A\ket{\hat{A}=A_x}=
A_x \ket{\hat{A}=A_x}
\;
\eeq
for $x\in Bool^\nb$.
Define the basis-changer unitary
operator $U_A$
by

\beq
\bra{y}U_A\ket{x}=
\av{y|\hat{A}=A_x}
\;
\eeq
 for $x,y\in Bool^\nb$.

Fig.\ref{fig-tom-pe}
includes at its bottom a time axis
marked with notches
for times 1 thru 4.
Let $\ket{\Psi_t}$
for $t\in Z_{1,4}$
denote
the quantum state
at those times.
Let us
represent
such states by sums
over 3 rows, where the
first row refers
to the $\nbj$ probe bits,
the second row refers
to the $\nb$ main bits,
and the final row
refers to the single
ancilla bit.

One has
\beqa
\ket{\Psi_1}&=&
\left[
\begin{array}{l}
\frac{1}{\sqrt{\nsj}}\sum_{j=0}^{\nsj-1}\ket{j}
\\
\sum_{x=0}^{\ns-1}
\ket{\hat{A}=A_x}
\bra{\hat{A}=A_x}V\ket{x_0}
\\
\ket{0}
\end{array}
\right.
\;,
\eeqa

\beqa
\ket{\Psi_2}&=&
\left[
\begin{array}{l}
\sum_{j'=0}^{\nsj-1}\ket{j'}\bra{j'}
\frac{1}{\sqrt{\nsj}}\sum_{j=0}^{\nsj-1}\ket{j}
\\
e^{i \hat{A}\Delta t j'}
\sum_{x=0}^{\ns-1}
\ket{\hat{A}=A_x}
\bra{\hat{A}=A_x} V\ket{x_0}
\\
\ket{0}
\end{array}
\right.
\\
&=&
\sum_{j=0}^{\nsj-1}
\sum_{x=0}^{\ns-1}
\left[
\begin{array}{l}
\ket{j}
\frac{e^{i A_{x}\Delta t j}}{\sqrt{\nsj}}
\\
\ket{\hat{A}=A_x}
\bra{\hat{A}=A_x} V\ket{x_0}
\\
\ket{0}
\end{array}
\right.
\;,
\eeqa

\beqa
\ket{\Psi_3}
&=&
\sum_{j=0}^{\nsj-1}
\sum_{x=0}^{\ns-1}
\left[
\begin{array}{l}
\sum_{j'=0}^{\nsj-1}
\ket{j'}\bra{j'}
U^\dagger_{FT}
\ket{j}
\frac{e^{i A_{x}\Delta t j}}{\sqrt{\nsj}}
\\
\ket{\hat{A}=A_x}
\bra{\hat{A}=A_x} V\ket{x_0}
\\
\ket{0}
\end{array}
\right.
\\
&=&
\sum_{x=0}^{\ns-1}
\left[
\begin{array}{l}
\sum_{j'=0}^{\nsj-1}
\ket{j'}
\sum_{j=0}^{\nsj-1}
\frac{e^{i (-\frac{2\pi j'}{\nsj}+A_{x}\Delta t) j}}{\nsj}
\\
\ket{\hat{A}=A_x}
\bra{\hat{A}=A_x} V\ket{x_0}
\\
\ket{0}
\end{array}
\right.
\\
&=&
\sum_{x=0}^{\ns-1}
\sum_{j'=0}^{\nsj-1}
\left[
\begin{array}{l}
\ket{j'}
\delta(\frac{2\pi j'}{\nsj},A_{x}\Delta t)
\\
\ket{\hat{A}=A_x}
\bra{\hat{A}=A_x} V\ket{x_0}
\\
\ket{0}
\end{array}
\right.
\\
&=&
\sum_{x=0}^{\ns-1}
\left[
\begin{array}{l}
\ket{j=A_x\frac{\Delta t \nsj}{2\pi}}
\\
\ket{\hat{A}=A_x}
\bra{\hat{A}=A_x} V\ket{x_0}
\\
\ket{0}
\end{array}
\right.
\;,
\eeqa
and

\beqa
\ket{\Psi_4}
&=&
\sum_{x=0}^{\ns-1}
\left[
\begin{array}{l}
\ket{j=A_x\frac{\Delta t \nsj}{2\pi}}
\\
\ket{\hat{A}=A_x}
\bra{\hat{A}=A_x} V\ket{x_0}
\\
\sqrt{\gamma f(A_x)}\ket{0} +
\sqrt{1-\gamma f(A_x)}\ket{1}
\end{array}
\right.
\;.
\label{eq-psi4}
\eeqa
Note that
we are
treating the quantity
$A_x\frac{\Delta t \nsj}{2\pi}$
as if it were an integer
in the range $Z_{0,\nsj-1}$.
This quantity is non-negative
because $A_x$ and $\Delta t$ are
non-negative by assumption.
$\Delta t$ can be chosen small enough
so that this quantity
is always smaller than $\nsj-1$. As
to whether the quantity
is an integer, it need
not be in general. When
it isn't, one has
to do a more careful
treatment (as done
in Ref.\cite{Har}).
By using the treatment
given above instead of a
more careful one like
the one of Ref.\cite{Har},
we are
introducing a small error
in what
we shall say next.

It follows from Eq.(\ref{eq-psi4})
that \footnote{Note
that
$\bra{\hat{A}=
A_\sam{x}{s}}=\bra{\sam{x}{s}}U^\dagger_A$
so if we were actually
going to observe the $\hat{A}$
of the main qubits, this
would require that we know and
apply $U^\dagger_A$
to them and then measure $\bra{\sam{x}{s}}$.
However, it turns out
that we don't need
to observe/measure the main qubits
to calculate $\check{P}_\rvb$, so we don't
need to know or apply $U^\dagger_A$
after all.}

\beq
\begin{array}{r}
\bra{j_f}
\\
\bra{\hat{A}=A_\sam{x}{s}}
\\
\bra{\sam{b}{s}}
\end{array}
\ket{\Psi_4}
={\scriptstyle
\delta^{j_f}_{
A_{\sam{x}{s}}\frac{\Delta t \nsj}{2\pi}
}
\left[
\bra{\hat{A}=A_\sam{x}{s}}
\sqrt{\gamma f(A)}
V\ket{x_0}\delta_{\sam{b}{s}}^0
+
\bra{\hat{A}=A_\sam{x}{s}}
\sqrt{1-\gamma f(A)}
V\ket{x_0}\delta_{\sam{b}{s}}^1
\right]}
\;,
\eeq
where $j_f\in Bool^\nbj$,
$\sam{x}{s}\in Bool^\nb$,
$\sam{b}{s}\in Bool$.
The index $s\in Z_{1,\nsam}$ labels $\nsam$
samples.
We can define an empirical
probability
distribution $\check{P}_{\rvx,\rvb}$
 by

\beq
\check{P}_{\rvx,\rvb}(x,b)
=
\frac{1}{\nsam}
\sum_{s=1}^{\nsam}
\delta_x^{\sam{x}{s}}
\delta_b^{\sam{b}{s}}
\;,
\eeq
where $x\in Z_{0,\ns-1}$
and $b\in Bool$.
Then

\beq
\check{P}_{\rvx,\rvb}(x,b)
\approx
\left|
\bra{\hat{A}=A_x}
\sqrt{\gamma f(A)}
V\ket{x_0}
\right|^2
\delta_{b}^0
+
\left|
\bra{\hat{A}=A_x}
\sqrt{1-\gamma f(A)}
V\ket{x_0}
\right|^2
\delta_{b}^1
\;.
\eeq
Therefore,

\beq
\check{P}_{\rvb}(b)=
\sum_{x=0}^{\ns-1}\check{P}_{\rvx,\rvb}(x,b)
=
\frac{1}{\nsam}
\sum_{s=1}^{\nsam}
\delta_b^{\sam{b}{s}}
\;,
\eeq
and

\beq
\check{P}_{\rvb}(b)\approx
\bra{x_0}V^\dagger
\gamma f(A)
V\ket{x_0}\delta_{b}^0
+
\bra{x_0}V^\dagger
[1-\gamma f(A)]
V\ket{x_0}\delta_{b}^1
\;.
\eeq
Finally, we conclude that

\beq
\bra{x_0}V^\dagger
f(A)
V\ket{x_0}
\approx
\frac{\check{P}_{\rvb}(0)}{\gamma}
=
\frac{1}{\gamma\nsam}
\sum_{s=1}^{\nsam}
\delta_0^{\sam{b}{s}}
\;.
\eeq
\qed

\section{Mean Values of Observables and \\
Boltzmann Partition Function}

In this section, we give
algorithms for calculating,
under various scenarios,
the expected value of an observable
$\Omega$ of a quantum system,
and the Boltzmann partition
function $Z=\sum_r e^{-\beta E_r}$
for an inverse temperature $\beta\in \RR^{\geq 0}$,
where $E_r$ are the eigenvalues
of the Hamiltonian of the quantum system.

Consider a quantum system with
density matrix $\rho$  and
 Hamiltonian $H$, where both operators
act on $\CC^\ns$.
(careful: $H$ is
also used for the single qubit Hadamard
transformation). Let

\beq
H\ket{\hat{H}=E_x}=
E_x \ket{\hat{H}=E_x}
\;
\eeq
for $x\in Bool^\nb$.
Define the basis-changer unitary operator $U_{H}$
by

\beq
\bra{y}U_H\ket{x}=
\av{y|\hat{H}=E_x}
\;
\eeq
for
$x,y\in Bool^\nb$.
Furthermore, consider a Hermitian operator
$\Omega$ acting on $\CC^\ns$
(what we call an
``observable" of
the quantum system) with

\beq
\Omega\ket{\hat{\Omega}=\Omega_x}=
\Omega_x \ket{\hat{\Omega}=\Omega_x}
\;
\eeq
and

\beq
\bra{y}U_\Omega\ket{x}=
\av{y|\hat{\Omega}=\Omega_x}
\;
\eeq
 for $x,y\in Bool^\nb$.

Some important scenarios
that we would like to consider are

\begin{enumerate}
\item[(a)] {\it Assume that
$\Omega_x$ for all $x\in Bool^\nb$
and $U_\Omega$
 are known.
Furthermore, assume that we know how to
p-compile
$U_\Omega$ and $e^{i\rho\Delta t}$
for any $\Delta t\geq 0$.
Then $\tr(\Omega\rho)$ can be calculated
as follows.}

In Claim \ref{cl-diag-mat-elem}, set
$V=U_\Omega$, $A=\rho$ and $f(\xi)=\xi$ for $\xi\in \RR^{\geq 0}$.
Thus

\beq
\mu(x)=
\bra{x} U_\Omega^\dagger
\rho
U_\Omega\ket{x}
\;\label{eq-px-case-a}
\eeq
can be calculated p-efficiently
for any $x\in Bool^\nb$.
Now a quantum sampling algorithm
like the quantum Gibbs sampling algorithm
of Ref.\cite{TucGibbsSam}
can be used to sample the probability
distribution
$P_\rvx(x)=\mu(x)$ given by Eq.(\ref{eq-px-case-a}).
If this yields the sample points $\sam{x}{s}$
for $s=1,2, \ldots \nsam$,
then

\beq
\tr(\Omega\rho)\approx
\frac{1}{\nsam}\sum_{s=1}^\nsam \Omega_\sam{x}{s}
\;.
\label{eq-omega-est}
\eeq

\item[(b)] {\it
Assume that
$\Omega_x$ for all $x\in Bool^\nb$
and $U_\Omega$
 are known.
Furthermore, assume that we know how to
p-compile
$U_\Omega$ and $e^{iH\Delta t}$
for any $\Delta t\geq 0$.
Let $\rho = \frac{e^{-\beta H}}{Z}$
where $\beta\in \RR^{\geq0}$ and
 $Z=\tr(e^{-\beta H})$.
Then $\tr(\Omega\rho)$ can be calculated
as follows.}

In Claim \ref{cl-diag-mat-elem}, set
$V=U_\Omega$, $A=H$ and $f(\xi)=e^{-\beta \xi}$ for $\xi\in \RR^{\geq 0}$.
(Assume that the
eigenvalues of $H$ are bounded below,
as is true for any physical Hamiltonian,
and, if necessary, add a constant to
$H$ so as to make
all its eigenvalues non-negative.)
Thus

\beq
\mu(x)=
\bra{x} U_\Omega^\dagger
e^{-\beta H}
U_\Omega\ket{x}
\;\label{eq-px-case-b}
\eeq
can be calculated p-efficiently
for any $x\in Bool^\nb$.
Now a quantum sampling algorithm
like the quantum Gibbs sampling algorithm
of Ref.\cite{TucGibbsSam}
can be used to sample
$\mu(x)$
given by Eq.(\ref{eq-px-case-b}).
(No need to know $\sum_x \mu(x)$ since
the sampling algorithm only uses probability ratios.)
If this yields the sample points $\sam{x}{s}$
for $s=1,2, \ldots \nsam$,
then an estimate of the expected
value of $\Omega$ is
again given by Eq.(\ref{eq-omega-est}).

\item[(c)]{\it
Assume that function $g:\RR^{\geq 0}\rarrow \RR^{\geq 0}$
is simple (that is,
that it can be calculated p-efficiently).
Assume that we know how to
p-compile
$e^{iH\Delta t}$
for any $\Delta t\geq 0$.
Let $\rho = \frac{e^{-\beta H}}{Z}$
where $\beta\in \RR^{\geq0}$ and
$Z=\tr(e^{-\beta H})$. Let
 $Z_g=\tr\{g(H)e^{-\beta H}\}$.
Then $Z_g$ and $\tr\{g(H)\rho\}$
can be calculated
as follows.}

In Claim \ref{cl-diag-mat-elem}, set
$V=1$, $A=H$ and
$f(\xi)=g(\xi)e^{-\beta \xi}$
for $\xi\in \RR^{\geq 0}$.
(If necessary,
add a constant to $H$ so as to make
all its eigenvalues non-negative.)
Thus

\beq
\mu(x)=
\bra{x}g(H)
e^{-\beta H}
\ket{x}
\;
\label{eq-px-case-c}
\eeq
can be calculated p-efficiently
for any $x\in Bool^\nb$.
Now a quantum sampling algorithm
like the quantum Gibbs sampling algorithm
of Ref.\cite{TucGibbsSam}
can be used to sample
$\mu(x)$
given by Eq.(\ref{eq-px-case-c}).
If this yields the sample points $\sam{x}{s}$
for $s=1,2, \ldots \nsam$,
then an estimate $\check{Z_g}$
of $Z_g$ can
be obtained as follows.

\beq
\check{P}_\rvx(x)=
\frac{1}{\nsam}
\sum_{s=1}^{\nsam}
\delta_x^\sam{x}{s}
\approx
\frac{\mu(x)}{Z_g}
\;,
\eeq

\beq
\check{Z_g}=
\frac{1}{\nsam}
\sum_{s=1}^{\nsam}
\frac{\mu(\sam{x}{s})}
{\check{P}_\rvx(\sam{x}{s})}
\approx Z_g
\;.
\eeq
To estimate $\tr\{g(H)\rho\}$, we can use

\beq
\frac{\check{Z_g}}{\check{Z_1}}
\approx \frac{Z_g}{Z_1}=
\tr\{g(H)\rho\}
\;.
\label{eq-est-tr-g-rho}
\eeq
Note that if $g:\RR^{\geq 0}\rarrow \RR$
instead of $g:\RR^{\geq 0}\rarrow \RR^{\geq 0}$,
one can set $g=g_+ - g_-$ where $g_\pm \geq 0$.
Then $Z_g = Z_{g_+}-Z_{g_-}$.
So we can estimate $Z_g$
by first estimating
$Z_{g_\pm}$. Since we can estimate
$Z_g$ and $Z_1$,
we can estimate $\tr\{g(H)\rho\}$
using Eq.(\ref{eq-est-tr-g-rho}).
\end{enumerate}

\section{Comparisons with the Joneses}

In this section,
we compare
the algorithms
of this paper with
algorithms in earlier papers
for calculating the same things (``our
competition").
This paper
has not addressed
the problem of finding
useful estimates and bounds
for the convergence rate
and error of its algorithms.
Such analysis will hopefully
be
done
in future papers
by me or others more capable.
Lacking this analysis,
 the comparisons in this
section should be taken as incomplete
and preliminary.

Suppose ${\cal V}$ is a Hilbert space and
$\rho$ is a density matrix acting
on $\CC^\ns$. Any pure state
$\ket{\Psi}\in \CC^\ns\otimes \calv$
such that $\tr_{\cal V}(\ket{\Psi}\bra{\Psi})=\rho$
is called a purification of $\rho$.

Given a probability distribution
$P_\rvx(x)$, where $x\in Z_{0,\ns-1}$,
it is convenient to define a
$\ket{\sqrt{P_\rvx}}$ state by

\beq
\ket{\sqrt{P_\rvx}} = \sum_{x=0}^{\ns-1}
\sqrt{P_\rvx(x)}\ket{x}
\;.
\eeq

Note that in the algorithms $(a),(b),(c)$,
finding the needed diagonal matrix elements
is p-efficient.
So the rate determining step of
these algorithms is the quantum sampling part,
which is not p-efficient.
However, if the quantum
sampling is done using the
algorithm of Ref.\cite{TucGibbsSam},
then the sampling converges
(to a given precision) in
$\calo(\frac{1}{\sqrt{\delta}})$ steps,
where $\delta$ is the distance
between the two largest eigenvalue
magnitudes of the underlying
Markov chain.

Note also that none of the
algorithms $(a),(b),(c)$
ever physically produces
a Boltzmann quantum state
(either $\rho = e^{-\beta H}/Z$
or a purification thereof).
They do produce $\ket{\sqrt{P_\rvx}}$
states.

Ref.\cite{Woc1}
by Wocjan et al.
presents an algorithm for
calculating the partition function
of a quantum system with
Hamiltonian $H$, but it
requires that the
eigenvalues of $H$ be known a priori
(as is the case when the system is classical).
The more recent Ref.\cite{Woc2}
by Poulin and Wocjan does
not require that the eigenvalues
of $H$ be known a priori.
However, the algorithm of Ref.\cite{Woc2}
physically creates a
sequence of Boltzmann quantum states (purifications
of $\rho = e^{-\beta H}/Z$),
each state characterized by
a different inverse temperature,
with the sequence
of inverse temperatures
increasing gradually,
according
to an ``annealing schedule",
from 0 towards
the target $\beta$.
Algorithm $(c)$
also calculates the
partition function of a
quantum system without
assuming that the eigenvalues
of $H$ are known a priori. But
algorithm $(c)$
never produces physically
any Boltzmann quantum state.
Nor does it introduce
error at every stage of
an annealing schedule (all
its calculations are done
at the target $\beta$).

Ref.\cite{Tem} by Temme et al.
gives an algorithm that
physically produces a
Boltzmann quantum state $\rho = e^{-\beta H}/Z$.
Instead of producing this
$\rho$ or a
purification of it, algorithm $(b)$
produces a $\ket{\sqrt{P_\rvx}}$ state.
If one wants to use the algorithm
of Temme et al.
to get $\tr(\Omega\rho)$,
one still must assume
the same things as $(b)$ above
(namely, that
$\Omega_x$ for all $x$
and $U_\Omega$
 are known, and that we know how to
p-compile
$U_\Omega$ and $e^{iH\Delta t}$).
If these assumptions are satisfied,
one can use the algorithm of Temme
et al. to produce
$U_\Omega^\dagger \rho U_\Omega$
physically, and then measure $\ket{x}\bra{x}$
on it. After obtaining $\nsam$ samples of $x$,
one can then use Eq.(\ref{eq-omega-est})
to estimate $\tr(\Omega\rho)$.
Thus, the Temme et al. algorithm
assumes the same things as $(b)$
to estimate $\tr(\Omega\rho)$.
A big advantage
of $(b)$ over Temme et al.
is that $(b)$ converges in
$\calo(\frac{1}{\sqrt{\delta}})$ steps,
whereas Temme et al.
converges, just like classical algorithms, in
$\calo(\frac{1}{\delta})$ steps.

\end{document}